\documentstyle[mprocl]{article}
\def\be{\begin{equation}}
\def\ee{\end{equation}}
\def\bea{\begin{eqnarray}}
\def\eea{\end{eqnarray}}

\newcommand{\venus}{ {\scriptscriptstyle +} \hspace{-1.5mm}^{\circ}}
\newcommand{\aeq}{&=&}

\newcommand{\itPi}{{\it \Pi}}

\newcommand{\itOmega}{{\it \Omega}}
\newcommand{\bra}{\langle}
\newcommand{\ket}{\rangle}
\newcommand{\dbra}{\bra \! \bra}
\newcommand{\dket}{\ket \! \ket}

\newcommand{\me}{\mbox{e}}

\newcommand{\strat}{\stackrel{\circ}{,}}

\begin{document}

\title{CAN "QUANTUMNESS" BE AN ORIGIN OF DISSIPATION?}

\author{T. ARIMITSU}

\address{Institute of Physics, University of Tsukuba,
Ibaraki,\\ 305-8571, Japan\\E-mail: arimitsu@cm.ph.tsukuba.ac.jp
}   


\maketitle\abstracts{ 
In their constructions of system of quantum stochastic 
differential equations, mathematicians and/or several
physicists interpret that the function of random force
operator is to preserve the canonical commutation relation
in time, i.e., to secure the unitarity of time evolution 
generator even for dissipative systems.
If this is the case, it means physically that the origin
of dissipation is attributed to quantum non-commutativity
({\it quantumness}).
The mechanism that the mathematician's approaches rest on will be 
investigated from the unified view point of 
Non-Equilibrium Thermo Field Dynamics (NETFD) 
which is a canonical operator 
formalism of quantum systems in far-from-equilibrium state
including the system of quantum {\it stochastic} equations.
}

\section{Introduction}

There are arguments 
\cite{Streater,Hasegawa,Hudson,Hudson85,Parth,Parth-text} that 
the function of random force operator is to preserve 
the canonical commutation relation
in time.
The contents of issue are the following.
The time-evolution of free Boson operators is given by
$
da(t)/dt = -i\omega a(t)
$,
$
da^\dagger(t)/dt = i\omega a^\dagger(t)
$.
The canonical commutation relation 
$
[a,\ a^\dagger]=1
$,
at time $t=0$ preserves in time, i.e.,
$
[a(t),\ a^\dagger(t)]=1
$.
If a relaxation is introduced simply by
\be
da(t)/dt = -i\omega a(t) -\kappa a(t),\quad
da^\dagger(t)/dt = i\omega a^\dagger(t)
-\kappa a^\dagger(t),
\ee
the canonical commutation relation decays as
$
[a(t),\ a^\dagger(t)]=\me^{-2\kappa t}
$.
This inconvenience is secured by introducing 
random operators $F(t)$ and $F^\dagger(t)$ which are assumed to
satisfy 
$
[a,\ F^\dagger(t)] =0
$,
$
[a^\dagger,\ F(t)] =0
$,
etc.\ for $t>0$.
The solutions of Langevin equations\footnote{
These stochastic differential equations should be interpreted as
those of the Stratonovich type~\cite{Strat}, since we perform
calculations as if they were ordinary differential equations.
}
\be
da(t)/dt = -i\omega a(t) -\kappa a(t)+ F(t), \quad
\label{math Lan a}\\
da^\dagger(t)/dt = i\omega a^\dagger(t)
-\kappa a^\dagger(t)+ F^\dagger(t),
\label{math Lan a dagger}
\ee
are given by
$
a(t) = a\me^{-i\omega t -\kappa t} 
+ \int_0^t dt' F(t') \me^{-i\omega(t-t') +\kappa (t+t')}
$,
$
a^\dagger(t) = a^\dagger \me^{i\omega t -\kappa t} 
+ \int_0^t dt' F^\dagger (t') \me^{i\omega(t-t') +\kappa (t+t')}
$.
Then, one knows that the canonical commutation relation
preserves in time if the condition
\be
1=[a(t),\ a^\dagger(t)] = \me^{-2\kappa t} \left\{
1 + \int_0^t dt_1 \int_0^t dt_2
 [F(t_1),\ F^\dagger(t_2)]
\me^{i\omega (t_1-t_2) +\kappa (t_1t_2)} \right\}
\ee
is satisfied. It is realized by the commutation relation among
the random force operators:
\be
[F(t),\ F^\dagger(t')] = 2\kappa \delta(t-t').
\label{random force commutator classical}
\ee

The above argument seems to show us that the function of 
random force operators is to recover the unitarity of 
time-evolution generator rather than to represent dissipative
thermal effects.
If it is correct, doesn't it mean, physically, that the origin
of dissipation can be attributed to quantum non-commutativity
({\it quantumness})?
There are, however, physical systems described by 
classical mechanics, or quantum systems with commutative 
random force operators.

In this paper, with the help of the framework of
Non-Equilibrium Thermo Field Dynamics 
(NETFD)~\cite{netfd1,Arimitsu1991,hydro,Naoko,Zubarev memorial,Saito97,Imagire98,Jp-It02}, 
we will investigate the above argument 
in a systematic manner by means of 
martingale operator paying attention to 
the {\it non-commutativity} among random force operators.
In section \ref{classical}, the structure of the system of 
stochastic differential equations in classical mechanics is
reviewed.
In section \ref{system q stoch diff eq}, the system of
quantum stochastic differential equations within NETFD
is introduced.
In section \ref{sys in rotating wave approx}, 
Boson system will be treated, which has a linear dissipative 
coupling with environment system within the rotating wave 
approximation. The mathematician's arguments will be
studied from the unified viewpoint based on 
the canonical operator formalism of NETFD by changing 
the intensity of non-commutativity parameter $\lambda$
among a martingale operator.
In section \ref{sys with commutative random operators}, 
Boson system having a linear dissipative 
interaction between environment {\it without} the rotating wave
approximation will be investigated with the help of NETFD.
This is the case where the system has commutative random 
force operators.
Section \ref{remarks} will be devoted to some remarks.

\section{System of Stochastic Differential Equations
in Classical Mechanics}
\label{classical}

\subsection{Stochastic Liouville Equation}

We will show the structure of the system of classical 
stochastic differential equations~\cite{kubo text} starting
with the stochastic Liouville equation 
\be
d f(u,t) = \Omega(u,t)dt \circ f(u,t),
\label{c-stoch Liouville}
\ee
of the Stratonovich type with
$
\Omega(u,t)dt = - (\partial/\partial u) du
$
where the flow $du$ in the velocity space is 
defined by
$
du = -\gamma u dt + m^{-1} dR(t)
\label{c-flow}
$.
Here, the circle $\circ$ represents 
the Stratonovich stochastic product~\cite{Strat}
which is defined in appendix A together with
the definition of the Ito stochastic product~\cite{Ito}.
The increment of random force $dR(t)$ is 
a Gaussian white stochastic process defined by
the fluctuation-dissipation theorem of the second kind:
\be
\bra dR(t) \ket = 0,\quad \bra dR(t) dR(t) \ket
= 2m \gamma T dt,
\label{FD2}
\ee
where $\gamma$ ($>0$) is a relaxation constant, and
$T$ a temperature of the environment represented by 
the random force $dR(t)$.
Note that within the Stratonovich calculus $dR(t)$ and
$f(u,t)$ are {\it not} stochastically independent:
$
\bra dR(t) \circ f(u,t) \ket \neq 0
$.
The average $\bra \cdots \ket$ is taken over 
all the possibility of the stochastic process $\{ dR(t) \}$.

By making use of the relation between 
the Ito and Stratonovich products
(\ref{connect-2}), the stochastic Liouville equation 
(\ref{c-stoch Liouville}) can be rewritten as the one of 
the Ito type:
\be
d f(u,t) = \itOmega(u,t)dt f(u,t),
\label{c-stoch Liouville Ito}
\ee
with
$
\itOmega(u,t)dt = - (\partial/\partial u) du
$
where $du$ is the flow in the Ito calculus given by
$
du = -\gamma \left( u + m^{-1}T (\partial/\partial u)
\right) dt + m^{-1} dR(t)
\label{c-flow Ito}
$.
Note that there appears temperature $T$ in the flow, and 
that within the Ito calculus $dR(t)$ and
$f(u,t)$ are stochastically independent:
$
\bra dR(t) f(u,t) \ket = 0
\label{orthogonal property}
$.

The initial condition for the stochastic distribution function
$f(u,t)$ is given by
$
f(u,0) = P(u,0)
$,
where $P(u,t)$ is the velocity distribution function defined
below in subsection \ref{P}.

Note that the stochastic distribution function conserves 
its probability within the relevant velocity phase-space:
$
\int du\ f(u,t) = 1
\label{conservation of probability}
$

\subsection{Langevin Equation}
The system described by the stochastic Liouville equation
(\ref{c-stoch Liouville}) can be treated by the Langevin equation:
\be
du(t) = -\gamma u(t) dt + m^{-1} dR(t).
\label{c-Langevin eq}
\ee
This Stratonovich type stochastic differential equation 
does {\em not} contain the diffusion term, which is very much 
related to the way how physicists originally introduced 
the Langevin equation.\footnote{
The Langevin equation was introduced by
adding a random force term, such as $m^{-1}dR(t)/dt$, to
a macroscopic phenomenological equation, for example, like
$
du(t)/dt = -\gamma u(t)
$.
}

It is worthwhile to note here that one could have introduced
the Langevin equation within the Ito calculus of the form
\be
du(t) = -\gamma [ u(t) + m^{-1}T 
(\delta/\delta u(t))] dt + m^{-1} dR(t),
\label{c-Langevin eq Ito}
\ee
which has a term with the functional derivative operator
$\delta/\delta u(t)$.
In the system of quantum stochastic differential equations 
within NETFD, the Langevin equation of the Ito type, 
such as (\ref{c-Langevin eq Ito}), can be introduced 
on the equal footing as the one of the Stratonovich type
(\ref{c-Langevin eq}), although this was not the original 
motivation for the invention of NETFD 
(see section \ref{system q stoch diff eq}).

\subsection{Fokker-Planck Equation}
In precise, the stochastic distribution function is given by
$
f(u,t) = f(\itOmega(u,t), P(u,0))
$.
Taking the random average $\bra \cdots \ket$,
we have an ordinary velocity distribution function 
$
P(u,t) = \bra f(\itOmega(u,t),P(u,0)) \ket
\label{P}
$
which satisfies the Fokker-Planck equation
\be
\partial P(u,t)/\partial t = (\partial/\partial u)
\gamma [u + m^{-1}T (\partial/\partial u) ]
P(u,t).
\label{c-F-P eq}
\ee
This can be derived most conveniently from 
the stochastic Liouville equation
(\ref{c-stoch Liouville Ito}) of the Ito type
because of the orthogonal property mentioned 
in subsection \ref{orthogonal property}.

The fluctuation-dissipation theorem (\ref{FD2}) 
of the second kind is introduced in order that 
the stochastic Liouville equation (\ref{c-stoch Liouville}) and 
the Langevin equation (\ref{c-Langevin eq}) are 
consistent with the Fokker-Planck equation (\ref{c-F-P eq}).

\section{System of Quantum Stochastic Differential Equations}
\label{system q stoch diff eq}

\subsection{Non-Equilibrium Thermo Field Dynamics}

In order to treat dissipative quantum systems dynamically, 
we constructed the framework of NETFD 
\cite{netfd1,Arimitsu1991,hydro,Naoko,Zubarev memorial,Saito97,Imagire98,Jp-It02}.
It is a {\it canonical operator formalism} of 
quantum systems in far-from-equili\-brium state 
which enables us to treat dissipative quantum systems 
by a method similar to the usual quantum field theory that
accommodates the concept of the dual structure in the
interpretation of nature, i.e.\ in terms of 
the {\it operator algebra} and the {\it representation space}.
In NETFD, the time evolution of the vacuum 
is realized by a condensation of 
$\gamma^{\venus} \tilde{\gamma}^{\venus}$-pairs into vacuum,
and that the amount how many pairs are condensed is described 
by the one-particle distribution function $n(t)$ whose 
time-dependence is given by corresponding kinetic equation 
(see appendix B).

We further succeeded to construct a unified framework of 
the {\it canonical operator formalism} for quantum 
{\it sto\-chas\-tic} differential equations with the help of 
NETFD. 
To the author's knowledge, it was not realized, until
the formalism of NETFD had been constructed, that one can
put all the stochastic differential equations 
for quantum systems into a unified method of canonical operator
formalism; 
the stochastic Liouville equation~\cite{kubo text} and the Langevin equation
within NETFD are, respectively, equivalent to the Schr\"odinger equation 
and the Heisenberg equation in quantum mechanics.
These stochastic equations are consistent with 
the quantum master equation which can be derived by taking 
random average of the stochastic Liouville equation.

\subsection{Quantum Stochastic Liouville Equation}

Let us start the consideration with 
the stochastic Liouville equation of the Ito type:
\be
d \vert 0_f(t) \ket = -i \hat{{\cal H}}_{f,t} dt\ 
\vert 0_f(t) \ket.
\label{Ito-Stoch Liou}
\ee
The generator $\hat{V}_{f}(t)$, defined by
$
\vert 0_f(t) \ket = \hat{V}_{f}(t) \vert 0 \ket
$,
satisfies 
$
 d \hat{V}_f(t) = -i \hat{{\cal H}}_{f,t} dt\ \hat{V}_f(t)
 	\label{eq:S-f-Ito}
$
with $\hat{V}_f(0)=1$. 
The stochastic hat-Hamiltonian $\hat{{\cal H}}_{f,t} dt$ is 
a tildian operator satisfying
$
(i \hat{{\cal H}}_{f,t} dt )^{\sim} 
= i \hat{{\cal H}}_{f,t} dt
$.
Any operator $A$ of NETFD is accompanied by its partner (tilde)
operator $\tilde{A}$, which enables us treat non-equili\-brium 
and dissipative systems by the method similar to 
usual quantum mechanics and/or quantum field theory. 
Here, the {\it tilde conjugation} $\sim$ is defined by
$
(A_1A_2)^{\sim} = \tilde{A}_1\tilde{A}_2
$,
$
(c_1A_1+c_2A_2)^{\sim} = c^{*}_1\tilde{A}_1+c^{*}_2\tilde{A}_2
$,
$
(\tilde{A})^{\sim} = A
$, and
$
(A^{\dagger})^{\sim} = \tilde{A}^{\dagger}
$
with $A$'s and $c$'s being operators and c-numbers, respectively.
The thermal ket-vacuum is tilde invariant:
$
\vert 0_f(t) \ket^{\sim} = \vert 0_f(t) \ket
$.

From the knowledge of the stochastic integral,
we know that the required form of the hat-Hamiltonian should be 
\be
\hat{{\cal H}}_{f,t} dt= \hat{H} dt\ + :d\hat{M}_t:
\label{H-ft Ito}
\ee
where $\hat{H}$ is given by 
$
\hat{H} = \hat{H}_{S} + i \hat{\itPi} 
	\label{Pi-1}
$
with
$
\hat{H}_S = H_S -\tilde{H}_S
\label{H-s}
$, and
$
\hat{\itPi} = \hat{\itPi}_R + \hat{\itPi}_D
\label{Pi}
$
where $\hat{\itPi}_R$ and $\hat{\itPi}_D$ are, respectively, 
the {\it relaxational} and the {\it diffusive} parts of 
the damping operator $\hat{\itPi}$.
The martingale $d\hat{M}_t$ is the term containing the operators 
representing the quantum Brownian motion
$dB_t$, $d\tilde{B}^\dagger_t$ and their tilde conjugates, and 
satisfies 
$
\bra \vert d\hat{M}_t \vert \ket = 0
$.
The symbol $:d\hat{M}_t:$ indicates to take the normal ordering
with respect to the annihilation and the creation operators both
in the relevant and the irrelevant systems 
(see (\ref{martingale with lambda})).

The operators of the quantum Brownian motion are introduced 
in appendix C, and satisfy 
the {\it weak} relations:
\bea
dB^\dagger_t\ dB_t \aeq \bar{n} dt,
\quad
dB_t\ dB^\dagger_t 
= \left( \bar{n}+1 \right) dt, 
\label{q Brown correlation}\\
d\tilde{B}_t\ dB_t \aeq \bar{n} dt,
\quad
d\tilde{B}^\dagger_t\ dB^\dagger_t 
= \left( \bar{n}+1 \right) dt,
\eea
and their tilde conjugates, with $\bar{n}$ being 
the Planck distribution function given in 
appendix B.
$\bra \vert$ and $\vert \ket$ are the vacuum states representing the 
quantum Brownian motion. They are tilde invariant:
$
\bra \vert^{\sim} = \bra \vert
$,
$
\vert \ket^{\sim} = \vert \ket.
$
It is assumed that, at $t=0$, a
relevant system starts to contact with the irrelevant
system representing the stochastic process included in the
martingale $d\hat{M}_t$.\footnote{
Within the formalism, the random force operators $dB_t$ 
and $dB^\dagger_t$ are assumed to 
commute with any relevant system operator $A$ in the 
Schr\"odinger representation: $[A,\ dB_t]= [A,\ dB^\dagger_t]
=0$ for $t \geq 0$.
}

\subsection{Quantum Langevin Equation}

The dynamical quantity $A(t)$ of the relevant system is defined by
the operator in the Heisenberg representation:
$
A(t) = \hat{V}_f^{-1}(t)\ A\ \hat{V}_f(t)
$
where $\hat{V}_f^{-1}(t)$ satisfies
$
d\hat{V}_f^{-1}(t) = \hat{V}_f^{-1}(t)\ i \hat{\cal H}_{f,t}^{-} dt
$
with
$
\hat{\cal H}_{f,t}^{-} dt = \hat{\cal H}_{f,t} dt
+ i d\hat{M}_t\ d\hat{M}_t
\label{hat-H minus}
$.

In NETFD, the Heisenberg equation for $A(t)$ within 
the Ito calculus is the quantum Langevin equation 
of the form
\be
dA(t) = i [ \hat{{\cal H}}_f(t) dt,\ A(t) ] 
	- d'\hat{M}(t)\  [ d'\hat{M}(t),\ A(t) ],
\label{Ito-Lan}
\ee
with
$
\hat{{\cal H}}_f(t) dt = \hat{V}_f^{-1}(t)\ 
 \hat{{\cal H}}_{f,t} dt\ \hat{V}_f(t)
$, and
\be
d'\hat{M}(t) = \hat{V}_f^{-1}(t)\ d\hat{M}_t\ \hat{V}_f(t).
\label{W-H}
\ee
Since $A(t)$ is an arbitrary observable operator in 
the relevant system, (\ref{Ito-Lan}) can be the Ito's formula 
generalized to quantum systems.

Applying the bra-vacuum $\dbra 1 \vert = \bra \vert \bra 1 \vert$ 
to (\ref{Ito-Lan}) from the left, 
we obtain the Langevin equation for the bra-vector $\dbra 1 \vert A(t)$
in the form
\be
d \dbra 1\vert A(t) = i \dbra 1\vert [H_S(t),\ A(t)] dt
+ \dbra 1 \vert A(t) \hat{\itPi}(t) dt
-i \dbra 1 \vert A(t)\ d'\hat{M}(t).
\label{Langevin for vector}
\ee
In the derivation, use had been made of the properties
$
\bra 1 \vert \tilde{A}^\dagger(t) = \bra 1 \vert A(t)
$,
$
\bra \vert d'\tilde{B}^\dagger(t) = \bra \vert d'B(t)
$, and
$
\dbra 1 \vert d'\hat{M}(t) = 0
$.

\subsection{Quantum Master Equation}

Taking the random average by applying the bra-vacuum 
$\bra \vert$ of the irrelevant sub-system 
to the stochastic Liouville equation
(\ref{Ito-Stoch Liou}), we can obtain the quantum master equation as
\be
(\partial/\partial t) \vert 0(t) \ket =
- i \hat{H} \vert 0(t) \ket,
\label{F-P}
\ee
with
$
\hat{H} dt = \bra  \vert \hat{\cal H}_{f,t} dt \vert \ket
$
and
$
\vert 0(t) \ket = \bra \vert 0_f(t) 
\ket.
$

\subsection{Stratonovich-Type Stochastic Equations}
By making use of the relation between the Ito and 
Stratonovich stochastic calculuses, 
we can rewrite the Ito stochastic Liouville equation 
(\ref{Ito-Stoch Liou}) and
the Ito Langevin equation (\ref{Ito-Lan}) into the Stratonovich ones,
respectively, i.e.,
\be
d \vert 0_f(t) \ket = -i \hat{H}_{f,t} dt \circ \vert 0_f(t) \ket,
\ee
with
$
\hat{H}_{f,t} dt = \hat{H}_S dt + i (
\hat{\itPi} dt + \frac12 d\hat{M}_t d\hat{M}_t ) 
+ d\hat{M}_t
\label{Liouville hat-H Stratonovich}
$, and
\be
dA(t) = i [\hat{H}_f(t) dt \strat A(t)],
\label{Strat-Lan}
\ee
with
$
\hat{H}_f(t) dt = \hat{H}_S(t) dt + i (
\hat{\itPi}(t) dt + \frac{1}{2} d'\hat{M}(t) d'\hat{M}(t)
) + :d'\hat{M}(t):
\label{Langevin hat-H Stratonovich}
$.

\subsection{Fluctuation-Dissipation Relation}

The fluctuation-dissipation theorem of the second kind 
for the multiple of martingales, $d\hat{M}_t\ d\hat{M}_t$, 
is determined by the criterion that there is no diffusive
term comes out in the terms
$
\hat{\itPi} dt + \frac12 d\hat{M}_t d\hat{M}_t
$
appeared in $\hat{H}_{f,t} dt$ in subsection \ref{Liouville hat-H Stratonovich}:
\be
d\hat{M}_t \ d\hat{M}_t = - 2 \hat{\itPi}_D dt.
\label{q F-D}
\ee
The origin of this criterion is attributed to the way how 
the Langevin equation was introduced in physics, as explained
before, i.e., relaxation term and random force term were introduced 
in mechanical equation within the Stratonovich calculus. 
Therefore, there is no dissipative terms in stochastic 
equations of the Stratonovich type.
We adopted this criterion in quantum cases.

The operator relation (\ref{q F-D}) may be called 
a generalized fluctuation-dissipation theorem of the second kind, 
which should be interpreted within the weak relation.

\section{A System in the Rotating Wave Approximation}
\label{sys in rotating wave approx}

\subsection{Model}

We will apply the above formalism to the model of a harmonic oscillator 
embedded in an environment with temperature $T$.
The Hamiltonian $H_{S}$ of the relevant system is given by
$
 H_{S}=\omega a^\dagger a
\label{H_S}
$
where $a,\ a^\dagger$ and their tilde conjugates
are stochastic operators of the relevant system
satisfying the canonical commutation relation
$
[ a,\ a^\dagger ] =1
$, and
$
[ \tilde{a},\ \tilde{a}^\dagger ] =1
$.
The tilde and non-tilde operators are 
related with each other by the relation
$
\bra 1 \vert a^\dagger = \bra 1 \vert \tilde{a}
$
where $\bra 1 \vert$ is
the thermal bra-vacuum of the relevant system.

Since we are interested in the system in the rotating wave
approximation, we will confine ourselves to the case where
the stochastic hat-Hamiltonian $\hat{{\cal H}}_t$ 
is bi-linear in $a,\ a^\dagger,\ dB_t,\ dB^\dagger_t$ and 
their tilde conjugates, and is invariant 
under the phase transformation
$a \rightarrow a \me^{i \theta}$, and $dB_t \rightarrow dB_t\ 
\me^{i \theta}$. This gives us the system of linear-dissipative coupling.

Then, $\hat{\itPi}_R$ and $\hat{\itPi}_D$ consisting of $\hat{\itPi}$
introduced in subsection \ref{Pi} become
\be
\hat{\itPi}_R = - \kappa
	( \gamma^{\venus} \gamma_\nu
	+ \tilde{\gamma}^{\venus} \tilde{\gamma}_\nu ), 
\label{Pi_R}
\quad
\hat{\itPi}_D = 2 \kappa ( \bar{n} + \nu ) 
	\gamma^{\venus} \tilde{\gamma}^{\venus},
\label{Pi_R,D}
\ee
respectively, where we introduced a set of canonical stochastic operators
$
\gamma_\nu = \mu a + \nu \tilde{a}^\dagger
$,
$
\gamma^{\venus} = a^\dagger - \tilde{a}
$
with $\mu + \nu =1$, which satisfy the commutation relation 
$
[\gamma_\nu,\ \gamma^{\venus}] = 1
$.
The parameter $\nu$ (or $\mu$) is closely related to the ordering
of operators when they are mapped to c-number function space 
with the help of the coherent state representation 
\cite{Zubarev memorial}, i.e., $\nu = 1$ for the normal ordering,
$\nu = 0$ for the anti-normal ordering, and $\nu = 1/2$ for 
the Weyl ordering.
The new operators $\gamma^{\venus}$ and $\tilde{\gamma}^{\venus}$
annihilate the relevant bra-vacuum:
$
\bra 1 \vert \gamma^{\venus} = 0,\quad 
\bra 1 \vert \tilde{\gamma}^{\venus} = 0
$.

\subsection{Martingale Operator}

Let us adopt the martingale operator:
\be
:d\hat{M}_t:\ =\ :d\hat{M}_t^{(-)}:\ +\ \lambda :d\hat{M}_t^{(+)}:
\label{martingale with lambda}
\ee
with
$
:d\hat{M}_t^{(-)}:\ = i ( \gamma^{\venus} dW_t
	+  \tilde{\gamma}^{\venus} d\tilde{W}_t )
\label{martingale minus}
$ and
$
:d\hat{M}_t^{(+)}:\ = - i ( dW^{\venus}_t \gamma_\nu
+  d\tilde{W}^{\venus}_t \tilde{\gamma}_\nu )
\label{martingale plus}
$.
Here, the annihilation and the creation random force operators 
$dW_t$ and $dW^{\venus}_t$ are defined, respectively, by
$
dW_t = \sqrt{2\kappa} ( \mu dB_t + 
\nu d\tilde{B}^\dagger_t )
\label{W-F0}
$,
$
dW^{\venus}_t = \sqrt{2\kappa} (dB^\dagger_t - 
d\tilde{B}_t )
$.
The latter annihilates the bra-vacuum $\bra \vert$ of the
irrelevant system:
$
\bra \vert dW^{\venus}_t= 0
$,
$
\bra \vert 
d\tilde{W}^{\venus}_t= 0
$.
Note that the normal ordering $: \cdots :$ is defined 
with respect to $\gamma$'s and $dW$'s.

The real parameter $\lambda$ measures the degree of non-commutativity
among the martingale operators:
$
[\ :d\hat{M}_t^{(-)}: ,\ :d\hat{M}_t^{(+)}:\ ] = -2 \hat{\itPi}_R dt
\label{non-commutativity}
$.
In deriving this, we used the facts that
\be
dW_t\ d\tilde{W}_t = d\tilde{W}_t\ dW_t = 2\kappa 
\left(\bar{n} + \nu \right) dt,
\label{WW}
\quad
dW_t\ dW_t^{\venus} = d\tilde{W}_t\ d\tilde{W}_t^{\venus}
= 2\kappa dt,
\label{WW venus}
\ee
and that the other combinations are equal to zero.
Note that 
$
[dW_t,\ dW_t^{\venus}] = 2\kappa dt
$
should be compared with (\ref{random force commutator classical}).
There exist at least two physically attractive cases~\cite{Zubarev memorial,Jp-It02},
i.e., one is the case for $\lambda = 0$ giving non-Hermitian martingale:
\be
d\hat{M}_t = i\sqrt{2\kappa} \left[ \left(a^\dagger - \tilde{a} \right)
d\tilde{B}_t^\dagger + \mbox{t.c.} \right],
\ee
and the other for $\lambda = 1$ giving Hermitian martingale:
\be
d\hat{M}_t = i\sqrt{2\kappa} [ \left( a^\dagger dB_t 
- dB_t^\dagger a \right) + \mbox{t.c.} ],
\ee
where t.c.\ stands for tilde conjugation.
The former follows the characteristics of the classical 
sto\-chastic Liouville equation where 
the sto\-chastic distribution 
function satisfies the conservation of probability within
the phase-space of a relevant system (see section \ref{classical}).
Whereas the latter employed the characteristics of 
the Schr\"odinger equation where the norm of the 
stochastic wave function preserves itself. In this case,
the consistency with the structure of classical system is
destroyed~\cite{Zubarev memorial,Jp-It02}.

The fluctuation-dissipation theorem of the system is given by
\be
:d\hat{M}_t:\ :d\hat{M}_t: = - 2 ( \lambda \hat{\itPi}_R 
+ \hat{\itPi}_D ) dt,
\ee
where we used the relations
$
:d\hat{M}_t^{(-)}:\ :d\hat{M}_t^{(-)}: = -2\hat{\itPi}_D dt
$,
$
:d\hat{M}_t^{(-)}:\ :d\hat{M}_t^{(+)}: = -2\hat{\itPi}_R dt
$ and
$
:d\hat{M}_t^{(+)}:\ :d\hat{M}_t^{(+)}: = :d\hat{M}_t^{(+)}:\ 
:d\hat{M}_t^{(-)}: =0
$,
which can be derived by making use of 
(\ref{WW}).

The hat-Hamiltonians of the model are given by
\be
\hat{H}_{f,t} dt = \hat{H}_S dt + i(1-\lambda) \hat{\itPi}_R dt
+ d\hat{M}_t,
\ee
\be
\hat{{\cal H}}^-_{f,t} dt = \hat{H}_S dt +i (
(1 - 2\lambda ) \hat{\itPi}_R - \hat{\itPi}_D ) dt
+ d\hat{M}_t,
\ee
\be
\hat{H}_{f}(t) dt = \hat{H}_S(t) dt 
+ i(1-\lambda) \hat{\itPi}_R(t) dt
+ :d'\hat{M}(t):.
\ee

\subsection{Heisenberg Operators of the Quantum Brownian Motion}
The Heisenberg operators of the Quantum Brownian motion are defined by
\be
B(t) = \hat{V}_f^{-1}(t)\ B_t\ \hat{V}_f(t),\quad
B^\dagger(t) = \hat{V}_f^{-1}(t)\ B^\dagger_t\ \hat{V}_f(t),
\ee
and their tilde conjugates. Their derivatives 
$
d B^\#(t) = d ( \hat{V}_f^{-1}(t)\ B^\#_t\ \hat{V}_f(t) )
$,
(\# : nul, dagger and/or tilde)
with respect to time in the Ito calculus are given, respectively, by 
\bea
dB(t) \aeq dB_t + \sqrt{2\kappa} \left[
\left( 1-\lambda \right) \nu \left( \tilde{a}^\dagger(t) 
- a(t) \right) - \lambda a(t) \right] dt,
\label{dB(t)}\\
dB^\dagger(t) \aeq dB^\dagger_t - \sqrt{2\kappa} \left[
\left( 1-\lambda \right) \mu \left( a^\dagger(t) 
- \tilde{a}(t) \right) + \lambda a^\dagger(t) \right] dt,
\label{dB(t)-dagger}
\eea
and their tilde conjugates. Then, we have
\be
dW(t) = dW_t - \lambda 2\kappa \gamma_\nu(t) dt,
\label{Heisenberg W1}
\quad
dW^{\venus}(t) = dW_t^{\venus} - 2\kappa \gamma^{\venus}(t) dt.
\label{Heisenberg W2}
\ee

Since, by making use of (\ref{Heisenberg W1}), we see that 
\be
d\hat{M}(t) = d'\hat{M}(t) = i [ \gamma^{\venus}(t) dW_t
	+  \tilde{\gamma}^{\venus}(t) d\tilde{W}_t ]
 - i \lambda [ dW^{\venus}_t \gamma_\nu(t)
	+  d\tilde{W}^{\venus}_t \tilde{\gamma}_\nu(t) ],
\ee
we know that the martingale operator in the Heisenberg representation
keeps the property:
$
\bra \vert d\hat{M}(t) \vert \ket = 0
$.

\subsection{Quantum Langevin Equations}

The quantum Langevin equation is given by
\bea
dA(t) \aeq i [\hat{H}_S(t),\ A(t)] dt
	\nonumber\\
	& & + \kappa \{(1-2\lambda) ( \gamma^{\venus}(t)
	[ \gamma_\nu(t),\ A(t) ]
+ \tilde{\gamma}^{\venus}(t)
	[ \tilde{\gamma}_\nu(t),\ A(t) ]
	)
	\nonumber\\
	& & \quad 
	+ [ \gamma^{\venus}(t),\ A(t) ]
	\gamma_\nu(t)
 + [\tilde{\gamma}^{\venus}(t),\ A(t)]
	\tilde{\gamma}_\nu(t)
	\} dt \nonumber\\
&&	 +  2 \kappa (\bar{n} + \nu )
	[ \tilde{\gamma}^{\venus}(t),\ 
	[ \gamma^{\venus}(t),\ A(t) ] ] dt
	\nonumber\\
	& & - \{
	[ \gamma^{\venus}(t),\ A(t) ] dW_t
	+ [ \tilde{\gamma}^{\venus}(t),\ A(t) ] d\tilde{W}_t
	\}
	\nonumber\\
	& & \quad + \lambda \{
	dW^{\venus}_t [ \gamma_\nu(t),\ A(t) ]
+ d\tilde{W}^{\venus}_t [ \tilde{\gamma}_\nu(t),\ A(t) ]
	\}
	\label{In Ito Langevin}
\\
	\aeq i [\hat{H}_S(t),\ A(t)] dt
	\nonumber\\
	& & + \kappa \{\gamma^{\venus}(t)
	[ \gamma_\nu(t),\ A(t) ]
	+ \tilde{\gamma}^{\venus}(t)
	[ \tilde{\gamma}_\nu(t),\ A(t) ]
	 \nonumber\\
	& & \quad 
	+ (1-2\lambda) ( [ \gamma^{\venus}(t),\ A(t) ]
	\gamma_\nu(t)
+ [\tilde{\gamma}^{\venus}(t),\ A(t)]
	\tilde{\gamma}_\nu(t)
	)
	\} dt 
	\nonumber\\
	& & +  2 \kappa (\bar{n} + \nu )
	[ \tilde{\gamma}^{\venus}(t),\ 
	[ \gamma^{\venus}(t),\ A(t) ] ] dt
	\nonumber\\
	& & - \{
	[ \gamma^{\venus}(t),\ A(t) ] dW(t)
	+ [ \tilde{\gamma}^{\venus}(t),\ A(t) ] d\tilde{W}(t)
	\}
	\nonumber\\
	& & \quad + \lambda \{
	dW^{\venus}(t) [ \gamma_\nu(t),\ A(t) ]
 + d\tilde{W}^{\venus}(t) [ \tilde{\gamma}_\nu(t),\ A(t) ]
	\},
	\label{Out Ito Langevin}
\eea
with 
$
\hat{H}_S(t) = \hat{V}_f^{-1}(t) \hat{H}_S \hat{V}_f(t)
= H_S(t) - \tilde{H}_S(t).
$
Note that the Langevin equation is written by means of 
the quantum Brownian motion
in the Schr\"odinger (the interaction) representation 
(the input field~\cite{Gardiner}) in (\ref{In Ito Langevin}),
and by means of that in the Heisenberg representation 
(the output field~\cite{Gardiner}) in (\ref{Out Ito Langevin}).

The Langevin equation for the bra-vector state,
$\dbra 1 \vert A(t)$, reduces to 
\bea
d \dbra 1\vert A(t) \aeq i \dbra 1\vert [H_S(t),\ A(t)] dt
\nonumber\\
&& - \kappa \left\{ \dbra 1\vert [A(t),\ a^\dagger(t)] a(t) 
+ \dbra 1\vert a^\dagger(t) [a(t),\ A(t)] \right\} dt
\nonumber\\
&& + 2\kappa \bar{n} \dbra 1\vert [a(t),\ [A(t),\ a^\dagger(t)]] dt
\nonumber\\
&&+\dbra 1\vert [A(t),\ a^\dagger(t)] \sqrt{2\kappa}\ dB_t 
+ \dbra 1\vert \sqrt{2\kappa}\ dB^\dagger_t [a(t),\ A(t)]
\label{In Ito final}
\\
\aeq i \dbra 1\vert [H_S(t),\ A(t)] dt
\nonumber\\
&& - \kappa (1-2\lambda) \left\{ \dbra 1\vert [A(t),\ a^\dagger(t)]
a(t) 
+ \dbra 1\vert a^\dagger(t) [a(t),\ A(t)] \right\} dt
\nonumber\\
&& + 2\kappa \bar{n} \dbra 1\vert [a(t),\ [A(t),\ a^\dagger(t)]] dt
\nonumber\\
&&+\dbra 1\vert [A(t),\ a^\dagger(t)] \sqrt{2\kappa}\ dB(t) 
+ \dbra 1\vert \sqrt{2\kappa}\ dB^\dagger(t) [a(t),\ A(t)].
\label{Out Ito final}
\eea
The relation between the expression (\ref{In Ito final}) and 
(\ref{Out Ito final}) can be interpreted as follows.
Substituting the {\it solution} of the Heisenberg random force operators
(\ref{dB(t)}) and (\ref{dB(t)-dagger}) for $dB(t)$ and $dB^\dagger(t)$,
respectively, into (\ref{Out Ito final}), we obtain 
the quantum Langevin equation (\ref{In Ito final}) which does not 
depend on the non-commutativity parameter $\lambda$.

The Langevin equations for $a(t)$ and $a^\dagger(t)$ of the system reduce to
\bea
da(t) \aeq \left(-i\omega -\kappa \right) a(t)dt + dW_t
 -2(1-\lambda) \nu \kappa \left[\tilde{a}^\dagger(t) - a(t) \right] dt
-\lambda \nu d\tilde{W}_t^{\venus},
\label{Lan RW a}\\
da^\dagger(t) \aeq \left( i\omega -\kappa \right) a^\dagger(t) dt 
+ d\tilde{W}_t
 +2(1-\lambda) \mu \kappa \left[ a^\dagger(t) 
- \tilde{a}(t) \right] dt + \lambda \mu d W_t^{\venus}.
\label{Lan RW a dagger}
\eea
Note that the last two terms in the above equations disappear
when one applies $\dbra 1 \vert$ to them.
For $\lambda = 0$, (\ref{Lan RW a}) and (\ref{Lan RW a dagger}) 
become, respectively, to
\bea
d a(t) \aeq -i\omega a(t) dt - \kappa \tilde{a}^\dagger(t) dt + dW_t,
\\
d a^\dagger(t) \aeq i\omega a^\dagger(t) dt - \kappa \tilde{a}(t) dt
+ d\tilde{W}_t,
\eea
where we put $\mu = \nu =1/2$, for simplicity.
For $\lambda = 1$, we get 
\bea
d a(t) \aeq -i\omega a(t) dt - \kappa a(t) dt + \sqrt{2\kappa} dB_t,
\label{Lan a RW}\\
d a^\dagger(t) \aeq i\omega a^\dagger(t) dt - \kappa a^\dagger(t) dt
+ \sqrt{2\kappa} dB^\dagger_t,
\label{Lan a dagger RW}
\eea
which may correspond to (\ref{math Lan a}).
Applying $\dbra 1 \vert = \bra 1 \vert \bra \vert$ 
to (\ref{Lan RW a}) and (\ref{Lan RW a dagger}), we obtain,
for any values of $\lambda$, $\mu$ and $\nu$, the Langevin equations
of the vectors $\dbra 1 \vert a(t)$ and $\dbra 1 \vert a^\dagger(t)$
in the forms
\bea
d \dbra 1 \vert a(t) \aeq -i\omega \dbra 1 \vert a(t) dt 
- \kappa \dbra 1 \vert a(t) dt 
 + \sqrt{2\kappa} \dbra 1 \vert dB_t,
\\
d \dbra 1 \vert a^\dagger(t) \aeq i\omega \dbra 1 \vert a^\dagger(t) dt 
- \kappa \dbra 1 \vert a^\dagger(t) dt
 + \sqrt{2\kappa} \dbra 1 \vert dB^\dagger_t.
\eea
Note that these have the same structure as those in (\ref{math Lan a}).

\section{A System with Commutative Random Force Operators}
\label{sys with commutative random operators}

Let us investigate Boson system having $x$-$X$ type interaction
between environment, i.e., a system without the rotating wave
approximation. The Hamiltonian of 
a harmonic oscillator can be written in the form
$
H_S = :p^2/(2m) + m\omega^2 x^2/2 :
$
with 
$
x = \sqrt{1/2m\omega} (a + a^\dagger )
$,
$
p = -i \sqrt{m\omega/2} (a-a^\dagger )
$
where $x$ and $p$ satisfy the canonical commutation relation 
$
[x,\ p] = i
$.
The normal ordering $: \cdots :$, here, is taken with respect to
$a$ and $a^\dagger$.
The relaxational and the diffusive parts in 
$
\hat{\itPi} = \hat{\itPi}_R + \hat{\itPi}_D
$
are given, respectively, as
\be
\hat{\itPi}_R = -i\kappa ( x - \tilde{x} ) 
( p + \tilde{p} ), \quad
\hat{\itPi}_D = - 2\kappa m \omega 
(\bar{n} + 1/2 ) ( x - \tilde{x} )^2.
\ee

The martingale operator corresponding to $x$-$X$ type interaction 
may have the form
\be
d \hat{M}_t = 2\sqrt{\kappa m \omega}
( x dX_t - \tilde{x} d\tilde{X}_t ),
\label{martingale x-X}
\ee
with
$
dX_t = (dB_t + dB_t^\dagger)/\sqrt{2}
$
where $dB_t$ and $dB_t^\dagger$ are the quantum Brownian motion 
defined in appendix C.
Then, we have
$
dX_t \ dX_t = dX_t \ d\tilde{X}_t = (\bar{n} + 1/2
)dt
$
which gives us the fluctuation-dissipation theorem
\be
d \hat{M}_t \ d \hat{M}_t = - 2 \hat{\itPi}_D dt.
\label{F-D theorem x-X}
\ee
The form of the martingale (\ref{martingale x-X}) was adopted by
following the structure of microscopic interaction 
Hamiltonian of the $x$-$X$ type.

The stochastic hat-Hamiltonian $\hat{{\cal H}}_{f,t} dt$ for 
the stochastic Liouville equation (\ref{Ito-Stoch Liou}) of 
the Ito type is given by
\be
\hat{{\cal H}}_{f,t} dt = \hat{H}_S 
+ i (\hat{\itPi}_R + \hat{\itPi}_D ) + d \hat{M}_t.
\ee
Then, the stochastic hat-Hamiltonian of the Stratonovich type
becomes
\be
\hat{H}_{f,t} dt = \hat{H}_S dt + i \hat{\itPi}_R dt + d\hat{M}_t,
\ee
where one does not see $\hat{\itPi}_D$ thanks to 
the fluctuation-dissipation theorem (\ref{F-D theorem x-X}).
We can also check that
\be
\hat{{\cal H}}^-_{f,t} dt = \hat{H}_S dt +i (
\hat{\itPi}_R - \hat{\itPi}_D ) dt
+ d\hat{M}_t.
\ee

The Langevin equation has the forms 
\bea
dx(t) \aeq \frac{1}{m} p(t) dt + \kappa 
( x(t) - \tilde{x}(t) ) dt,
\label{Lan x}\\
dp(t) \aeq -m\omega^2 x(t) dt - \kappa 
(p(t) + \tilde{p}(t) ) dt \nonumber\\
&& + 4i\kappa m \omega (\bar{n} + 1/2 )
(x(t) - \tilde{x}(t) ) dt
 - 2\sqrt{\kappa m \omega}\ dX_t.
\label{Lan p}
\eea

Applying $\bra 1 \vert$ to (\ref{Lan x}) and (\ref{Lan p}),
we have the Langevin equation for $\bra 1 \vert x(t)$ 
and $\bra 1 \vert p(t)$ in the forms
\bea
d\bra 1 \vert x(t) \aeq m^{-1} \bra 1 \vert p(t) dt,
\label{Lan vec x}\\
d\bra 1 \vert p(t) \aeq -m\omega^2 \bra 1 \vert x(t) dt
- 2\kappa \bra 1 \vert p(t) dt
 - 2\sqrt{\kappa m \omega}\ \bra 1 \vert dX_t,
\label{Lan p vec}
\eea
respectively.
They can be written in terms of $a$ and $a^\dagger$ as
\bea
d\bra 1 \vert a(t) \aeq -i\omega \bra 1 \vert a(t) dt
-\kappa \bra 1 \vert \left[ a(t) - a^\dagger(t) \right] dt
-i \sqrt{\kappa} \bra 1 \vert \left( dB_t + dB_t^\dagger \right),
\label{Lan vec a}\\
d\bra 1 \vert a^\dagger(t) \aeq i\omega \bra 1 \vert a^\dagger(t) dt
-\kappa \bra 1 \vert \left[ a^\dagger(t) - a(t) \right] dt
-i \sqrt{\kappa} \bra 1 \vert \left( dB_t^\dagger + dB_t \right).
\eea
If we take the rotating wave approximation at this stage
the coefficients in front of the quantum Brownian motion
are not equal to those appeared in (\ref{Lan a RW}) 
and (\ref{Lan a dagger RW}).
It may indicate that a naive procedure of taking the rotating wave
approximation will not give us correct results.
It might also be related to the renormalization procedure
needed to derive stochastic differential equations
for the system with $x$-$X$ type interaction 
from a microscopic Heisenberg equation~\cite{Saito99}.

\section{Concluding Remarks}
\label{remarks}

We have revealed that the non-commutativity among
$d\hat{M}_t^{(-)}$ and $d\hat{M}_t^{(+)}$ appeared in 
the martingale operator of the model within the rotating wave
approximation affect the relaxation part of the stochastic 
hat-Hamiltonian. When the measure $\lambda$ of the non-commutativity
has the value $\lambda = 1$, the hat-Hamiltonian becomes 
Hermite, and therefore, it looks like being related to 
a microscopic description.
On the other hand, for $\lambda = 0$, the system of the quantum stochastic 
differential equations has the same structure as that of classical mechanics,
and it is related to a semi-macroscopic description.
As has been shown in this paper, Hermiticy of the hat-Hamiltonian
is realized thanks to the non-commutativity between 
$d\hat{M}_t^{(-)}$ and $d\hat{M}_t^{(+)}$.

Does this mean that the system with commutative martingale does not
have any microscopic realization? 
As an example of system with commutative martingale, we studied 
the system corresponding to the quantum Kramers equation which has
$x$-$X$ type interaction Hamiltonian between the relevant system 
and environment system.
Since, there is no non-commutative parts in the martingale operator,
this system cannot have an Hermitian stochastic hat-Hamiltonian.
The non-commutative parts appears when one takes the rotating wave 
approximation to the interaction Hamiltonian. 
Does dissipation originate in the {\it approximation} causing non-commutative
character in martingale? Can {\it quantumness}, appeared in this way,
be the origin of dissipation?
On the contrary, the following question arises naturally.
Is it always possible to put all random force operators 
to be commutative?

There are still a lot of problems to be resolved before we know
the origin of dissipation. However, with the help of NETFD,
we can see the problems from a unified viewpoint which may
provide us with good prospects for further developments.
Introducing the parameter $\lambda$ in the martingale term
as given by (\ref{martingale with lambda}), we can transform 
the equation to the non-Hermitian version by shifting 
$\lambda \rightarrow 0$ (see (\ref{Out Ito Langevin})).
In other words, it seems that the non-commutativity
is renormalized into the relaxational and diffusive terms.

Substituting the solution of the random force operators 
(\ref{dB(t)}) and (\ref{dB(t)-dagger}) 
in the Heisenberg representation (the output field) into 
(\ref{Out Ito Langevin}), 
we have the Langevin equation (\ref{In Ito Langevin}) 
expressed by means of those 
in the Schr\"odinger (or, more properly, the interaction) 
representation (the input field). 
Note that the Langevin equation (\ref{In Ito final}) for the bra-vector state
$\dbra 1 \vert A(t)$ does not depend on $\lambda$ when
it is represented by the random force operator in
the Schr\"odinger representation (the input field).

We are intensively investigating what is the physical meaning of 
the renormalization of non-commutativity by changing 
the parameter $\lambda$.
The relation between the present argument and the procedure
of the coarse graining is under investigation.
Related to the system with commutative random force
operators, a microscopic derivation of 
quantum stochastic equations corresponding to 
the quantum Kramers equation are in progress~\cite{Saito99}. 
There, an appropriate renormalization is required 
in accordance with the separation of two time-scales, i.e.,
microscopic and macroscopic time-scales.
Without the renormalization, one gets quantum stochastic 
equations in the rotating wave approximation, which do 
not correspond to the system described by 
the Kramers equation.
Including these studies, the further progress will be reported 
elsewhere.

\vspace*{-2pt}

\section*{Acknowledgments}
Somewhat preliminary contents of 
the paper was presented at the International Workshop,
{\it New Developments in Statistical Physics}, 
held in University of Tokyo in 1997,
celebrating Prof.~M.~Suzuki's sixtieth birthday 
at the occasion of his retirement from University of Tokyo.
The present version was mostly developed at 
Prigogine Center for Statistical Mechanics and
Complex Systems in the University of Texas at Austin 
in Autumn, 1999. The author would like to express 
his sincere thanks to all members of the center, especially
to Prof.~I.~Prigigine and Dr.~T.~Petrosky, for their
warm hospitality in the productive tense atmosphere.

\vspace*{-2pt}

\section*{Appendix A}

The definitions of the Ito~\cite{Ito} and the Stratonovich 
\cite{Strat} stochastic products are given, respectively, by
\be
X_t \ dY_t = X_t \left( Y_{t + dt} 
- Y_t \right),
	\label{def-Ito-1}
\quad
dX_t \ Y_t = \left( X_{t + dt} - X_t 
	\right) Y_t
	\label{def-Ito-2}
\ee
and
\be
X_t \circ dY_t = \frac{X_{t + dt} + X_t}{2}
	\left( Y_{t + dt} - Y_t \right),
	\label{def-Strat-1}
\quad
dX_t \circ Y_t = \left( X_{t + dt} - X_t 
	\right)
	\frac{Y_{t + dt} + Y_t}{2}
	\label{def-Strat-2}
\ee
for arbitrary stochastic operators $X_t$ and
$Y_t$.
From (\ref{def-Ito-1}) and (\ref{def-Strat-1}), we have the formulae
which connect the Ito and the Stratonovich products in the
differential form
\be
X_t \circ dY_t = X_t \ dY_t 
	+ (1/2) dX_t \ dY_t,
	\label{connect-1}\quad
dX_t \circ Y_t = dX_t \ Y_t
	+ (1/2) dX_t \ dY_t.
	\label{connect-2}
\ee

\section*{Appendix B}
The time-evolution of the thermal vacuum $\vert 0(t) \ket$, 
satisfying the quantum master equation (\ref{F-P})
with the hat-Hamiltonian for the semi-free system 
specified by $H_{\rm S}=\omega a^\dagger a$ and (\ref{Pi_R,D}), is given by 
\be
 \vert0(t)\ket \ = \exp \left\{ \left[n(t)-n(0) \right] \gamma^{\venus} 
 \tilde{\gamma}^{\venus} \right\} \vert0\ket,
\label{b3}
\ee
where the one-particle distribution function, 
$
n(t) = \dbra 1 \vert a^\dagger(t) a(t) \vert 0 \dket,
$
satisfies the kinetic (Boltzmann) equation of the model:
$
 dn(t)/dt = -2\kappa \left[ n(t) - \bar{n} \right]
$
with the Planck distribution function
$
\bar{n} = \left( \me^{\omega/T} -1 \right)^{-1}
\label{Planck distribution}
$.
Here, $T$ is the temperature of environment system.

\section*{Appendix C}

Let us introduce the annihilation and creation operators 
$b_t$, $b^\dagger_t$ and their tilde conjugates 
satisfying the canonical commutation relation:
\be
[ b_t,\ b^\dagger_{t'} ] = \delta(t - t'),\quad 
[ \tilde{b}_t,\ \tilde{b}^\dagger_{t'} ] = \delta(t - t').
\ee
The vacuums $( 0 \vert$ and $\vert 0 )$ are defined by 
$
b_t \vert 0 ) = 0
$,
$
\tilde{b}_t \vert 0 ) = 0
$ and
$
( 0 \vert b^\dagger_t = 0,\quad ( 0 \vert 
\tilde{b}^\dagger_t = 0
$.
The subscript or the argument $t$ represents time.

Introducing the operators
$
B_t = \int_0^{t-dt} dB_{t'} = \int_0^t dt'\ b_{t'}
$,
$
B^\dagger_t = \int_0^{t-dt} dB^\dagger_{t'} = \int_0^t dt'\ b^\dagger_{t'}
\label{B}
$
and their tilde conjugates for $t \geq 0$, we see that they satisfy
$
B(0) = 0
$, 
$
B^\dagger(0)=0
$, 
$
[B_s,\ B^\dagger_t ] = \mbox{min}(s,t)
\label{commutation B}
$,
and their tilde conjugates, and that 
they annihilate the vacuums $\vert 0 )$ and $( 0 \vert$:
$
dB_t \vert 0 ) = 0
$,
$d\tilde{B}_t \vert 0 ) = 0
$,
$
( 0 \vert d B^\dagger_t =0
$,
$
( 0 \vert d\tilde{B}^\dagger_t =0
$.
These operators represent the quantum Brownian motion.

Let us introduce a set of new operators by the relation
$
dC_t^\mu = {\cal B}^{\mu \nu} dB_t^\nu
$
with the Bogoliubov transformation defined by
\bea
 {\cal B}^{\mu\nu} = \left(
  \begin{array}{cc}
   1+ \bar{n} & -\bar{n} \\
   -1 & 1 \\
  \end{array}
 \right),
\label{B bar}
\eea
where $\bar{n}$ is the Planck distribution function.
We introduced the thermal doublet:
\be
dB_t^{\mu=1} = dB_t,\quad dB_t^{\mu=2} = d\tilde{B}^\dagger_t,
\quad
d\bar{B}_t^{\mu =1} = dB^\dagger_t,\quad d\bar{B}_t^{\mu = 2} = -
d\tilde{B}_t,
\ee
and the similar doublet notations for $dC_t^\mu$ and 
$d{\bar C}_t^\mu$.
The new operators annihilate the new vacuum $\bra \vert$ and 
$\vert \ket$:
$
dC_t \vert  \ket = 0
$,
$
d\tilde{C}_t \vert  \ket = 0
$,
$
\bra  \vert d C^\dagger_t =0,\quad \bra \vert 
d\tilde{C}^\dagger_t =0
\label{cal B bra}
$.

We will use the representation space constructed on
the vacuums $\bra \vert$ and $\vert \ket$.  
Then, we have, for example,
\be
\bra \vert dB_t \vert \ket = \bra \vert dB^\dagger_t \vert \ket = 0,
\quad
\bra \vert dB^\dagger_t dB_t \vert \ket = \bar{n} dt,
\quad
\bra \vert dB_t dB^\dagger_t \vert \ket 
= \left( \bar{n}+1 \right) dt.
\label{B-Bd}
\ee

\end{document}